\documentclass[fleqn,twoside]{article}
\usepackage{espcrc2}
\usepackage{graphicx}

\newcommand{\AmS}{{\protect\the\textfont2
  A\kern-.1667em\lower.5ex\hbox{M}\kern-.125emS}}

\makeatletter
\def\alt{\mathrel{\mathpalette\vereq<}}
\def\vereq#1#2{\lower3pt\vbox{\baselineskip1.5pt \lineskip1.5pt
\ialign{$\m@th#1\hfill##\hfil$\crcr#2\crcr\sim\crcr}}}
\def\agt{\mathrel{\mathpalette\vereq>}}
\makeatother

\title{\vbox to 0pt{{\ }\newline\vskip-3.5cm \noindent
\normalsize\rm Contribution to the Proceedings of the
Seventh International Workshop on Topics in Astroparticle 
and Underground Physcis (TAUP 2001), 
8--12 Sept.~2001, Laboratori Nazionali del Gran Sasso, Italy,
to be published in Nuclear Physics B (Proceedings Supplements).
\vfil}
\vskip -24pt
Physics with Supernovae}

\author{Georg G.~Raffelt\\
{\ }\\
Max-Planck-Institut f\"ur Physik
(Werner-Heisenberg-Institut), F\"ohringer Ring 6, 
80805 M\"unchen, Germany}

\begin{document}

\begin{abstract}
Core-collapse supernovae (SNe) are powerful neutrino sources and as
such important targets for the growing array of neutrino
observatories.  We review the current status of SN theory and the
expected characteristics of the neutrino signal.  After recalling what
we have learned from SN~1987A and general SN properties we review the
physics potential of a future galactic SN observation.
\vspace{1pc}
\end{abstract}

\maketitle


\section{\uppercase{Supernova Types and Rates}}

Supernovae are exploding
stars~\cite{Brown,Woosley:ta,Petschek,Burrows:mk,Cappellaro}.
However, there are two entirely different classes, both of which are
of current interest for astro-particle physics and cosmology.  One
physical class are the type~Ia supernova (SN) explosions. A SN~Ia is
thought to occur when a carbon-oxygen white dwarf accretes matter from
a companion star until it reaches its Chandrasekhar limit and begins
to collapse, thereby triggering a nuclear explosion, powered by the
fusion of carbon and oxygen to heavier nuclei.  SNe~Ia are
spectroscopically characterized by the absence of hydrogen and the
presence of silicon lines.  The explosion disrupts the progenitor
white dwarf entirely; what remains is an expanding nebula without a
central compact object. While the exact SN~Ia lightcurves depend on
some parameters, they are surprisingly reproducible and thus lend
themselves as cosmological standard candles.  The main astro-particle
interest in SNe~Ia is their potential to explore the space-time
geometry of the universe; the observed SN~Ia Hubble diagram suggests
the presence of ``dark energy'' or a cosmological
constant~\cite{Riess:1998cb,Perlmutter:1998np}.

\begin{table*}
\caption{Supernova rates in $h^2\,{\rm SNu}$ according 
to Refs.~\cite{Cappellaro,Cappellaro:1999qy}.}
\label{tab:SNrates}
\renewcommand{\arraystretch}{1.2} 
\begin{tabular}{lcccc}
\hline
Galaxy    & \multicolumn{4}{c}{Supernova type}\\
\cline{2-5}
type      &     Ia        &      Ib/c    &    II      &  All   \\
\hline
E--S0      &  $0.32\pm.11$ &    $<0.02$   &   $<0.04$  & $0.32\pm.11$\\
S0a--Sb    &  $0.32\pm.12$ & $0.20\pm.11$& $0.75\pm.34$& $1.28\pm.37$\\
Sbc--Sd    &  $0.37\pm.14$ & $0.25\pm.12$& $1.53\pm.62$& $2.15\pm.66$\\
All       &  $0.36\pm.11$ & $0.14\pm.07$& $0.71\pm.34$& $1.21\pm.36$\\
\hline
\end{tabular}
\end{table*}

The present lecture is exclusively about the other class of explosions
which mark the evolutionary end of massive stars ($M\agt 8\,M_\odot$).
Such stars have the usual onion structure with several burning shells,
an expanded envelope, and a degenerate iron core that is essentially
an iron white dwarf. The core mass grows by the nuclear burning at its
edge until it reaches the Chandrasekhar limit. The collapse can not
ignite nuclear fusion because iron is the most tightly bound
nucleus. Therefore, the collapse continues until the equation of state
stiffens by nucleon degeneracy pressure at about nuclear density
($3\times10^{14}~{\rm g~cm^{-3}}$).  At this ``bounce'' a shock wave
forms, moving outward and expelling the stellar mantle and
envelope. The explosion is a reversed implosion, the energy derives
from gravity, not from nuclear energy. Within the expanding nebula, a
compact object remains in the form of a neutron star or perhaps
sometimes a black hole. The kinetic energy of the explosion carries
about 1\% of the liberated gravitational binding energy of about
$3\times10^{53}~{\rm erg}$, 99\% going into neutrinos. This powerful
and detectable neutrino burst is the main astro-particle interest of
core-collapse SNe; the Ia explosions do not produce significant
neutrino emission.  In core-collapse SNe only $10^{-4}$ of the total
energy shows up as light, i.e.~about 1\% of the kinetic explosion
energy.  Core-collapse SNe are dimmer than SNe~Ia, and their
lightcurves are different from case to case, the details depending on
the structure of the progenitor star.  Core-collapse SNe are not
useful as standard candles.

If the progenitor star has retained a hydrogen envelope, hydrogen
lines will appear in the lightcurve, qualifying the SN
spectroscopically as type~II, while type~I are the ones without
hydrogen lines. If the star has lost its hydrogen envelope (all stars
suffer significant mass loss during their giant phase), but has
retained helium, the helium lines in the SN lightcurve make it a
type~Ib. Without hydrogen and helium lines it is of type~Ic, unless it
shows silicon lines, which characterize a type~Ia.  Confusingly the
spectroscopic types Ib, Ic and II form the physical class of
core-collapse SNe.

Table~\ref{tab:SNrates} gives the observed SN rates for different
galaxy types according to Refs.~\cite{Cappellaro,Cappellaro:1999qy},
some of them significantly smaller than the rates in an earlier
review~\cite{vandenBergh1991}.  The SN rate is expressed in the
``Supernova unit,'' defined as $1~{\rm SNu}=1~{\rm SN}$ per
$10^{10}\,L_{\odot,B}$ per 100~yrs where $L_{\odot,B}$ is the solar
luminosity in the blue spectral band.  Therefore, 1~SNu corresponds
roughly to 1~SN per galaxy per century. Moreover, $h$ is the Hubble
constant in units of $100~{\rm km~s^{-1}~Mpc^{-1}}$.  Early-type
galaxies, where little star formation takes place, do not host
core-collapse SNe as this type depends on the formation of massive
stars which are short-lived on cosmological scales.  About 2/3 of all
SNe are core collapse, and of those the vast majority is type~II
(hydrogen lines). On the other hand, because SNe~Ia are intrinsically
brighter, the majority of observed SNe are of that type. About 2000
SNe have been observed, but many have not been classified---for an
up-to-date catalogue see~\cite{Asiago}.

For the field of neutrino astronomy, the most crucial question is the
SN rate in our own Milky Way because even the largest foreseen
detectors will not reach beyond our galaxy and its satellites.  The
closest big galaxy, Andromeda (M31), is at a distance of about
0.7~Mpc. Even in a megatonne detector a SN in Andromeda would yield
only about 30 events. 

One approach to estimate the galactic SN rate is to apply the relevant
average rate of Table~\ref{tab:SNrates} to the Milky Way.  Assuming a
morphological type Sb--Sbc, a blue luminosity of
$2.3\times10^{10}\,L_{\odot,B}$, and a Hubble constant $h=0.75$ one
finds $2\pm1$ core-collapse SNe per century \cite{Cappellaro}, about a
factor of~2 smaller than the corresponding estimate
in~\cite{vandenBergh1991} or~\cite{Tammann:ev}. Note that the
morphological type of our galaxy is not well determined.

Another approach relies on the historical SN record, extrapolated to
the entire galaxy. (Because of obscuration by dust, only SNe out to a
few kpc have been observed.) The rate of core-collapse SNe is then
estimted to be 3--4 per century~\cite{Tammann:ev,Strom1994}, with a
large Poisson uncertainty from the small number of observed cases
(5~SNe during the second millenium).

Given the vagaries of small-number statistics, these estimates agree
with each other, and with circumstantial evidence such as the
estimated population of progenitor stars or the neutron-star formation
rate. Except for SN~1987A in the Large Magellanic Cloud, no neutrino
burst has been observed, even though large neutrino detectors have
been in operation continuously since the Baksan Scintillator Telescope
began operations in June 1980~\cite{Alekseev:dy}.  This
non-observation is in agreement with the estimated SN rate and
suggests that stellar collapse events without SN explosions are not
frequent relative to normal SNe.


\begin{figure}[b]
\includegraphics[width=\hsize]{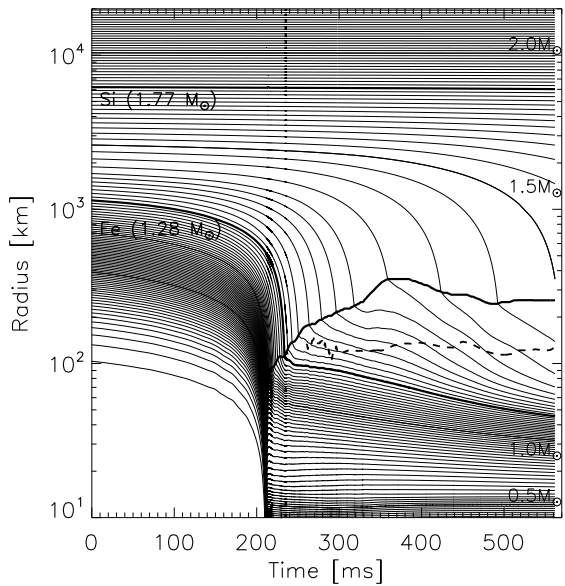}
\caption{Trajectories of mass shells, spaced in steps of
$0.02\,M_{\odot}$, in the spherically symmetric calculation of Rampp
and Janka~\cite{Rampp:2000ws}.  Also shown are the outer boundaries of
the iron core (at $1.28\, M_{\odot}$) and of the silicon shell (at
$1.77\,M_{\odot}$).  The shock is formed at $211~{\rm ms}$, its
position is also marked by a bold line.  The dashed curve shows the
position of the gain radius.  (Figure from Ref.~\cite{Rampp:2000ws}
with permission.)}
\label{fig:1}
\end{figure}

\section{\uppercase{Core-Collapse Supernova\\ Explosion Mechanism}}

The bounce-and-shock explosion scenario of core-collapse
SNe~\cite{Brown,Woosley:ta,Petschek,Burrows:mk} is essentially a
hydrodynamic phenomenon---see, for example, Ref.~\cite{Font} which
includes very intuitive animations. However, realistic numerical
simulations have difficulties exploding for a physical reason.  The
shock wave at core bounce forms {\it within\/} the iron core.  As it
moves outward energy is dissipated by the dissociation of iron.  The
nuclear binding energy of $0.1~M_\odot$ iron is about
$1.7\times10^{51}~{\rm erg}$ and thus of the same order as the SN
explosion energy. Therefore, the shock wave stalls without driving off
the stellar mantle and envelope. This behavior is illustrated in
Fig.~\ref{fig:1} which represents a state-of-the-art spherically
symmetric collapse calculation~\cite{Rampp:2000ws}. The figure shows
the trajectories (radial position vs.\ time) of selected mass shells,
and also shows the boundaries of the iron core and silicon shell as
well as the shock trajectory. The shock wave stagnates at about 200~km
while mass accretion continues---mass shells continue to cross the
shock position. The shock wave never ``takes off'' to explode the
star. Similar state-of-the-art results are reported by the
Oakridge group~\cite{Mezzacappa:2000jb}.

The standard scenario of SN explosions holds that the stagnating shock
will be ``re-juvenated'' by energy deposition so that enough pressure
builds up behind the shock to set it back into motion.  This ``delayed
explosion scenario'' was first proposed in the early 1980s by Bethe
and Wilson~\cite{Bethe:1984ux}.  One source of energy deposition
behind the shock wave is energy absorption from the nearly freely
streaming neutrinos which originate from the neutrino sphere near the
neutron-star surface.  The required conditions for a successful shock
revival have been studied numerically and analytically---see
Ref.~\cite{Janka:2000bt} for details and references.  Continued mass
accretion and convection below the shock wave also deposit energy and
thus contribute to the shock revival.

The main recent progress in numerical SN calculation has been the
implementation of efficient Boltzmann solvers so that an exact
neutrino transport scheme can be self-consistently coupled with the
hydrodynamic
evolution~\cite{Rampp:2000ws,Mezzacappa:2000jb,Burrows:tj}.  Such
state-of-the-art spherically symmetric calculations do not lead to
successful explosions. However, these calculations are not
self-consistent in that the regions below the shock wave are
convectively unstable. Likewise, convection may arise in the neutron
star below the neutrino sphere.  Forthcoming calculations will reveal
if convection, perhaps coupled with more accurate neutrino interaction
rates, will lead to successful explosions.

The Livermore group does obtain robust
explosions~\cite{Totani:1997vj}.  In their spherically symmetric
calculations they include a mixing-length treatment of ``neutron
finger convection,'' thereby enhancing the early neutrino luminosity
and thus the energy deposition behind the
shock~\cite{Wilson:rx}. Their results agree with the findings of other
groups that diffusive neutrino transport alone is not enough to
trigger the explosion.

The delayed explosion scenario may involve new particles or new
interactions. Of course, too much energy deposition in the SN mantle
would make the explosions too energetic, providing limits on radiative
neutrino decays~\cite{Falk:kf}. On the other hand, new particles could
transfer additional energy from the inner core to the shock wave and
thus trigger the explosion~\cite{Schramm:1981mk,Berezhiani:1999qh}.
An intriguing scenario involving resonant neutrino flavor oscillations
would have required mass differences much larger than indicated by
current oscillation experiments and thus is no longer
viable~\cite{Fuller}.

It is not known at present if the standard delayed explosion scenario
is the correct picture, or if new physical ingredients beyond the
self-consistent inclusion of neutrino transport and convection are
needed. Even if robust explosions are obtained in future 2- and
3-dimensional calculations, the long-standing problem of the large
neutron-star velocities remains unresolved---for a recent review
see~\cite{Lai}.

The high-statistics neutrino light curve from a future galactic SN in
a large neutrino detector would allow one to observe directly the
collapse dynamics. For example, the early accretion-powered neutrino
emission could be clearly distinguished from the subsequent
neutron-star cooling phase~\cite{Totani:1997vj}.  One of the most
energetic astrophysical phenomena would be caught in the act, allowing
one to unravel the underlying physics.


\section{\uppercase{Expected Neutrino Signal}}

\begin{figure}
\includegraphics[width=\hsize]{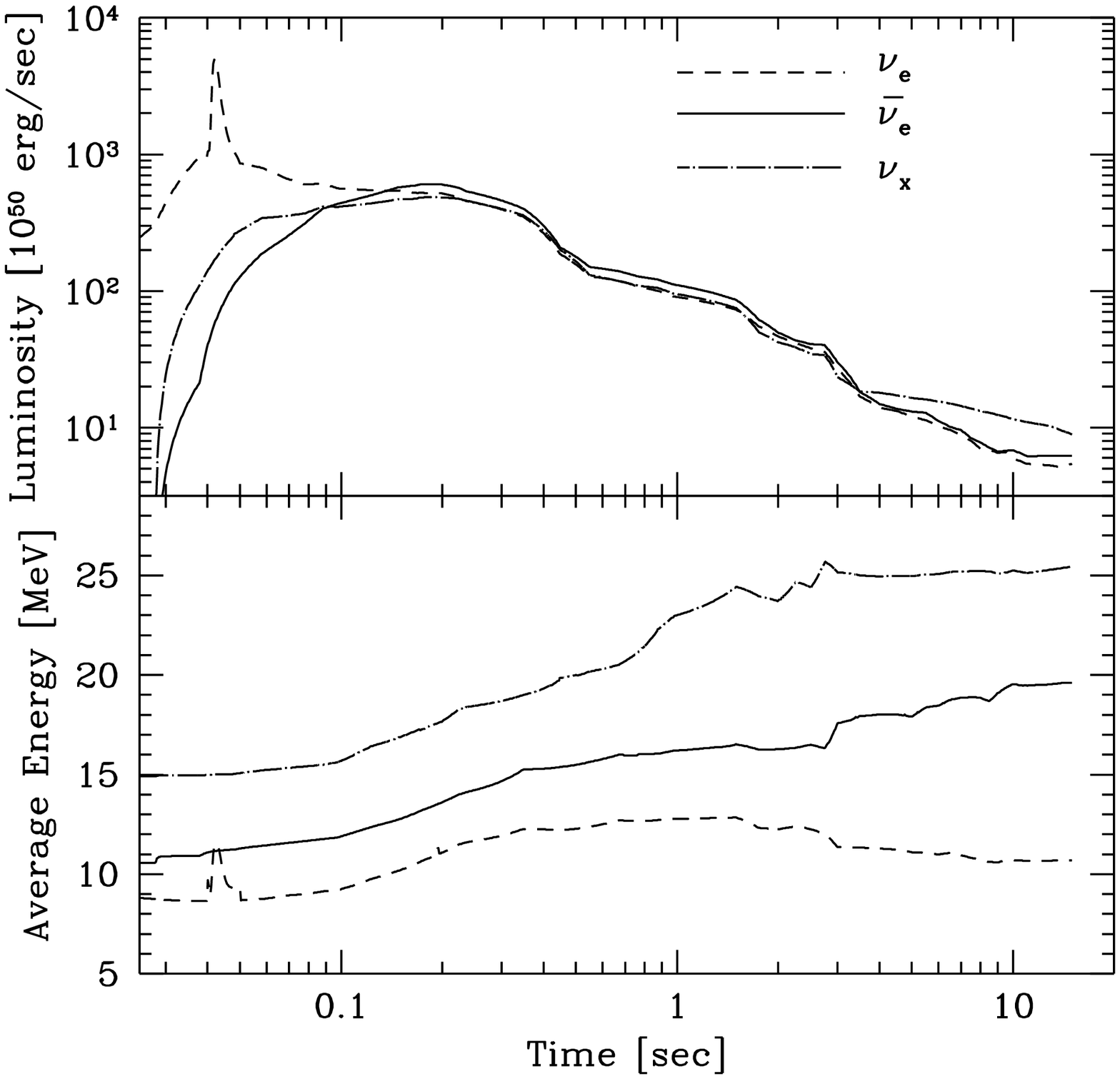}
\includegraphics[width=\hsize]{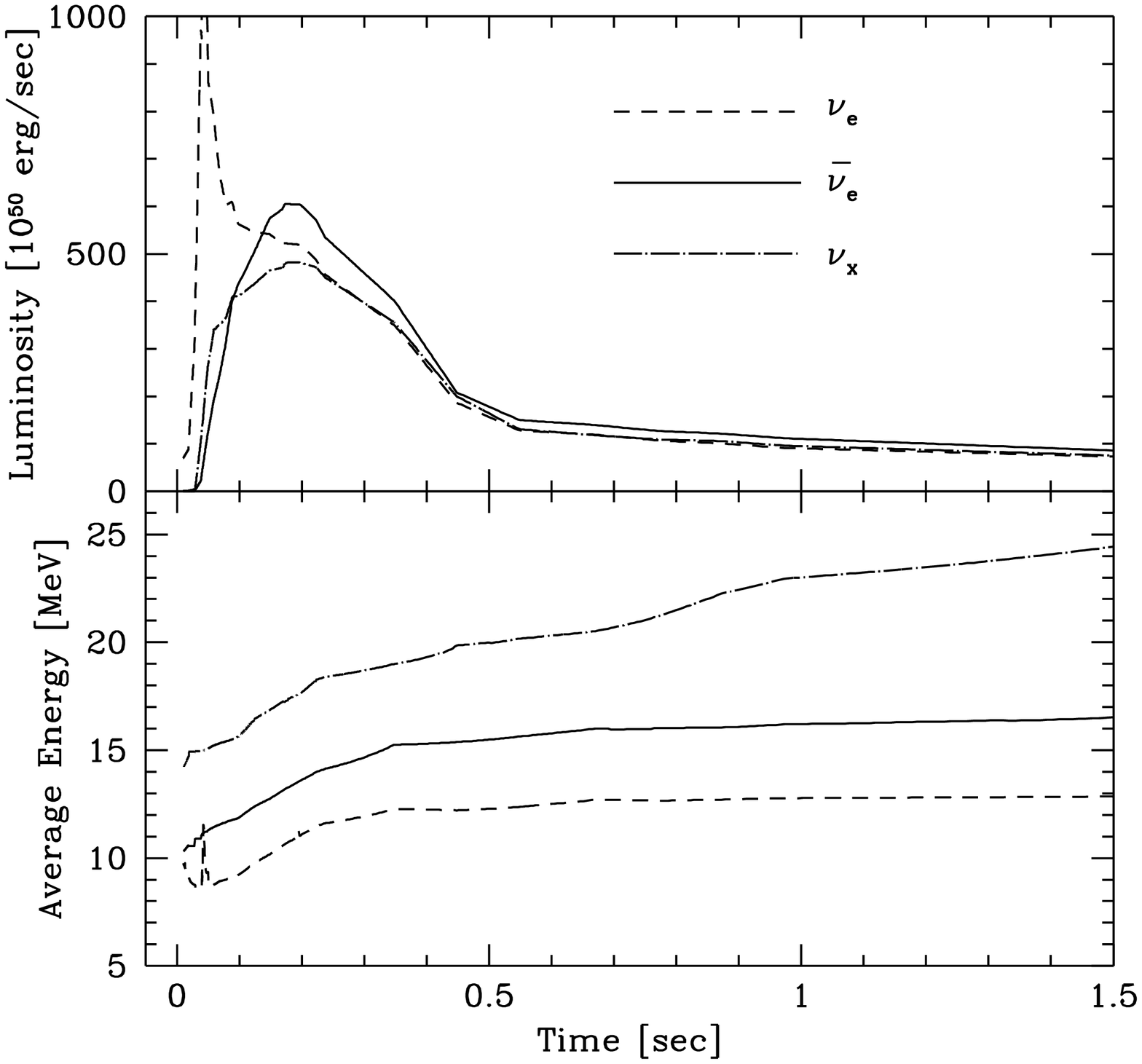}
\caption{Neutrino luminosities and average energies in a SN collapse
and explosion simulation with the Livermore code.  The $\nu_x$ line
represents each of $\nu_\mu$, $\bar\nu_\mu$, $\nu_\tau$ and
$\bar\nu_\tau$.  Upper panel: Logarithmic luminosity and time scales.
Upper panel: Linear scales.  (Figures from Ref.~\cite{Totani:1997vj}
with permission.)}
\label{fig:2}
\end{figure}

The expected neutrino fluxes and spectra are illustrated by the
numerical results shown in Fig.~\ref{fig:2}. The $\nu_e$ lightcurve
shows a conspicuous spike early on, representing the prompt neutrino
burst which occurs when the shock wave reaches the region of neutrino
trapping in the iron core. The dissociation of iron allows for the
quick neutronization of a layer of the proto neutron star. Of course,
most of the lepton number remains trapped and slowly escapes by
neutrino diffusion.

The subsequent broad shoulder up to about 500~ms, best visible in the
lower panel with linear scales, represents the accretion phase where
material keeps falling in and powers the neutrino emission.  After
this phase the shock wave has driven off the stellar mantle.  The
subsequent long and flat tail represents the neutron star cooling by
neutrino emission.

The duration of the accretion phase depends on how long it takes to
revive the shock wave. In the Livermore simulation, an explosion is
obtained by a phenomenological treatment of neutron-finger convection
which boosts the early neutrino luminosity~\cite{Wilson:rx}.  In the
absence of a confirmed robust explosion mechanism the exact duration
of the accretion phase is not known---other simulations do not obtain
explosions and thus do not get beyond the accretion phase.

After about 100~ms, the neutrino luminosities are virtually equal in
each flavor. This equipartition of the emitted energy is almost
perfect in this simulation. In the recent Oakridge simulation
\cite{Mezzacappa:2000jb}, which includes a state-of-the-art Boltzmann
solver, the equipartition is also nearly perfect between $\nu_e$ and
$\bar\nu_e$, but the $\nu_x$ luminosity is less than 1/2 after 50~ms
out to 600~ms when this simulation terminates.  Therefore,
``equipartition'' probably should be taken to mean ``equal to within
about a factor of two.''

The neutrino average energies obey the well-known hierarchy $\langle
E_{\nu_e}\rangle <\langle E_{\bar\nu_e}\rangle <\langle
E_{\nu_x}\rangle$ which is explained by the different trapping
processes, $\beta$~processes for the electron flavor and elastic
scattering on nucleons for the rest. Therefore, the different flavors
originate in layers with different temperatures.  A physical
understanding of the neutrino spectra can be developed without
large-scale numerical simulations~\cite{Raffelt:2001kv}.  While the
flavor hierarchy of average energies appears to be generic, the
differences are likely smaller than previously thought after all
relevant processes have been included, notably nucleon bremsstrahlung
and energy transfer by recoils~\cite{Raffelt:2001kv,Janka:1995ir}.
However, no state-of-the-art numerical simulation yet exists that
includes all of the relevant microphysics.

In all numerical simulations the $\nu_\mu$, $\bar\nu_\mu$, $\nu_\tau$,
and $\bar\nu_\tau$ are treated equally. However, the transport of
neutrinos and anti-neutrinos is different even for the heavy flavors
because the cross section for $\nu N\to N\nu$ is different from
$\bar\nu N\to N\bar\nu$ because of weak
magnetism~\cite{Horowitz:2001xf}. Moreover, the presence of muons is
not entirely negligible, at least in the deep interior of the SN core
so that muonic beta reactions are also possible.  While this may not
affect the spectra formation near the neutron star atmosphere, it is
not assured that the $\mu$- and $\tau$-flavored neutrino spetra are
the same. In principle, then, neutrino transport in a SN core involves
six different neutrino degrees of freedom. However, muons are not
included in the available equations of state, and treating all flavors
differently enhances the numerical CPU requirements.

The average neutrino energies increase for the first few seconds. This
is a generic effect because the neutrino-emitting regions heat up by
accretion and by the contraction of the neutron star.  Of course,
eventually the average energies must decrease when the neutron star
cools.

Numerical neutrino light curves can be compared with the SN~1987A data
where the measured energies are found to be ``too low.''  For example,
the numerical simulation of Fig.~\ref{fig:2} yields time-integrated
values $\langle E_{\nu_e}\rangle\approx13~{\rm MeV}$, $\langle
E_{\bar\nu_e}\rangle\approx16~{\rm MeV}$, and $\langle
E_{\nu_x}\rangle\approx23~{\rm MeV}$.  On the other hand, the data
imply $\langle E_{\bar\nu_e}\rangle=7.5~{\rm MeV}$ at Kamiokande and
11.1~MeV at IMB~\cite{Jegerlehner:1996kx}.  Even the 95\% confidence
range for Kamiokande implies $\langle E_{\bar\nu_e}\rangle<12~{\rm
MeV}$.  Flavor oscillations would increase the expected energies and
thus enhance the discrepancy~\cite{Jegerlehner:1996kx}.  It has
remained unclear if these and other anomalies of the SN~1987A neutrino
signal should be blamed on small-number statistics, or point to a
serious problem with the SN models or the detectors.


\section{\uppercase{Observing a Future Galactic Supernova}}

\begin{figure}
\includegraphics[width=\hsize]{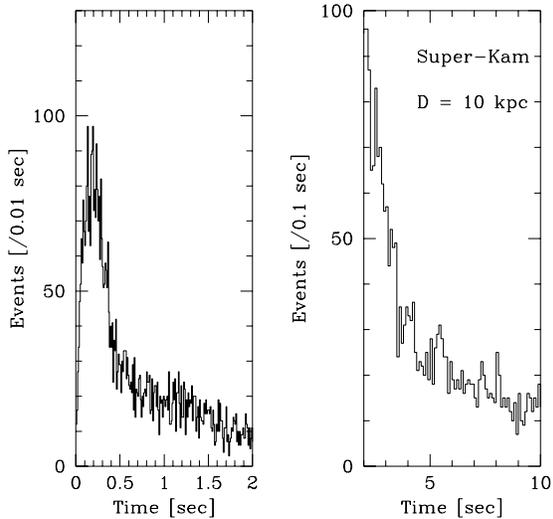}
\caption{Simulated Super-Kamiokande neutrino light curve for
a galactic SN at 10~kpc. (Figures from Ref.~\cite{Totani:1997vj}
with permission.)}
\label{fig:3}
\end{figure}

Detectors for measuring the neutrino signal from a galactic SN have
almost continuously operated since 1980 when the Baksan Scintillator
Telescope (BST) took up operation.  For a galactic SN at a distance of
10~kpc with neutrino fluxes and spectra roughly like those of
Fig.~\ref{fig:2}, BST would register about 70~events.  The neutrinos
from SN~1987A in the Large Magellanic Cloud at a distance of 50~kpc
were actually measured in Kamiokande~\cite{Hirata:ad},
IMB~\cite{Bionta:1987qt}, and BST~\cite{Alekseev:gp} with a few events
each. Today, much larger detectors are available, although BST keeps
running.  Super-Kamiokande would measure about 8000 events for a SN at
10~kpc. A simulated light curve based on the SN model of
Fig.~\ref{fig:2} is shown in Fig.~\ref{fig:3}.

Super-Kamiokande is not operational at the time of this writing
because of the destructive accident during re-filling on 12~November
2001. The exact capabilities for SN neutrino detection after repair
are not known at present, but hopefully will not be dramatically
worse.  In the following, all statements concerning the
Super-Kamiokande capabilities rely on the pre-accident literature.

Fortunately, there are other large detectors available. The Sudbury
Neutrino Observatory (SNO) would register about 800 events from our
fiducial SN~\cite{Virtue:2001mz}, where for now we ignore flavor
oscillations. The Large Volume Detector (LVD) in the Gran Sasso
Laboratory is a scintillation detector that would register about
400~events~\cite{Aglietta:2001jf}. A similar number of events would be
expected in the KamLAND scintillation reactor neutrino experiment
which recently began taking data~\cite{Piepke:tg}.  The Borexino solar
neutrino experiment, that will soon be ready, is smaller and would
register about 100~events~\cite{Cadonati:2000kq}. The AMANDA South
Pole neutrino telescope also works as a SN neutrino detector in that
the correlated noise of all photomultipliers caused by the Cherenkov
light of the SN neutrinos produces a significant signal, especially
when AMANDA is enlarged to the cubic-kilometer
IceCube~\cite{Ahrens:2001tz}.

The dominant signal is usually the charged-current reaction $\bar\nu_e
p\to n e^+$.  SNO has a unique $\nu_e$ detection capability from the
charged-current deuterium dissociation $\nu_ed\to
ppe^-$. Neutral-current reactions which are sensitive to all flavors
include elastic scattering on electrons, the deuterium dissociation
$\nu d\to np\nu$ in SNO, the excitation of $^{16}\rm O$ in water
Cherenkov detectors, and the corresponding excitation of $^{12}\rm C$
in scintillation detectors, notably in LVD and KamLAND, where the
$\gamma$-rays from the subsequent de-excitation can be measured.
Another recent suggestion is the elastic scattering on protons which
can cause a measurable signal in low-threshold scintillation
detectors~\cite{Beacom2001,Vogel:2001yi}.

Specific neutral-current detectors for SN neutrinos have been proposed
on the basis of the reaction $\nu+(A,Z)\to (A-1,Z)+n+\nu$ where the
neutron will be measured.  For example, lead or iron could be used as
targets in the proposed OMNIS detector~\cite{Smith:td,Boyd2001}.  This
sort of detector would be complementary to Super-Kamiokande and SNO in
that it is primarily sensitive to the heavy-flavor neutrinos.

At present one debates the possibility of building even larger
detectors for the purpose of precision neutrino long-baseline
oscillation experiments, for proton decay, and high-statistics solar,
atmospheric and SN neutrino detection. A typical size could be
a megatonne of water or scintillator.  This option is discussed under
the name of Hyper-Kamiokande in Japan~\cite{Shiozawa2001}, under UNO
in the US~\cite{Vagins2001}, and is also debated in
Europe~\cite{Frejus2002}.  Such a detector could produce as many as
$10^5$ events from our fiducial SN at 10~kpc.

The operation of large neutrino detectors is motivated by many physics
goals so that it is not unrealistic to expect that another few decades
will be covered by neutrino observatories sensitive to a galactic
SN. Therefore, even though the galactic SN rate is low, the chance of
observing one within a few decades is not small so that it is
worthwhile to discuss the possible benefits from such an observation.

Arguably the most important gain would be the direct observation of
stellar collapse where a high-statistics neutrino observation would
map out the dynamics of a cataclysmic astrophysical event that could
never be observed directly in any other way. Whether or not numerical
SN simulations will soon converge on a theoretical standard model for
the collapse and explosion mechanism, the importance of its
independent verification or falsification by a detailed neutrino light
curve can not be overstated.

Another benefit is the possibility of an early warning for the
occurrence of a SN because the neutrino signal precedes the optical
explosion by several hours.  This project has been taken up by the
Supernova Early Warning System (SNEWS), a network of detectors with SN
neutrino capabilities~\cite{SNEWS}.  Unfortunately, the triangulation
of the SN by the arrival time at various detectors is relatively poor.
However, the electron recoil signal in Super-Kamiokande can locate the
SN within a circle of radius $7^\circ$--$8^\circ$ in the
sky~\cite{Beacom:1998fj,Ando:2001zi}.  A future megatonne detector
probably could do much better.


\section{\uppercase{Flavor Oscillations}}

Neutrino oscillations are now firmly established so that the SN
neutrino fluxes and spectra expected in a detector can be very
different from those emitted at the source. This is especially true if
the solar neutrino problem is solved by the large-mixing angle (LMA)
case, which is presently favored, and which can be confirmed or
refuted by the KamLAND experiment in the immediate
future~\cite{Piepke:tg}.  The relevant mass difference of $\Delta
m_{12}^2=1$--$10\times10^{-5}~{\rm eV}^2$ implies that matter effects
are important in the SN and also in the Earth if the neutrinos happen
to enter the detector ``from below.''  The large ``solar'' mixing
angle $\theta_{12}$ implies that oscillations will be important in
both the $\nu_e$ and the $\bar\nu_e$ channel.

If the LMA case obtains, it is unavoidable that oscillation effects
influence the SN~1987A signal interpretation, and that the detectors
saw different spectra due to different Earth-crossing segments of the
neutrino paths~\cite{Jegerlehner:1996kx,Smirnov:ku,Lunardini:2000sw,%
Kachelriess:2001sg}.  While this effect can make the measurements
slightly more consistent with each other, the unexpectedly soft
neutrino energies become even more worrisome.

\begin{figure}
\includegraphics[width=\hsize]{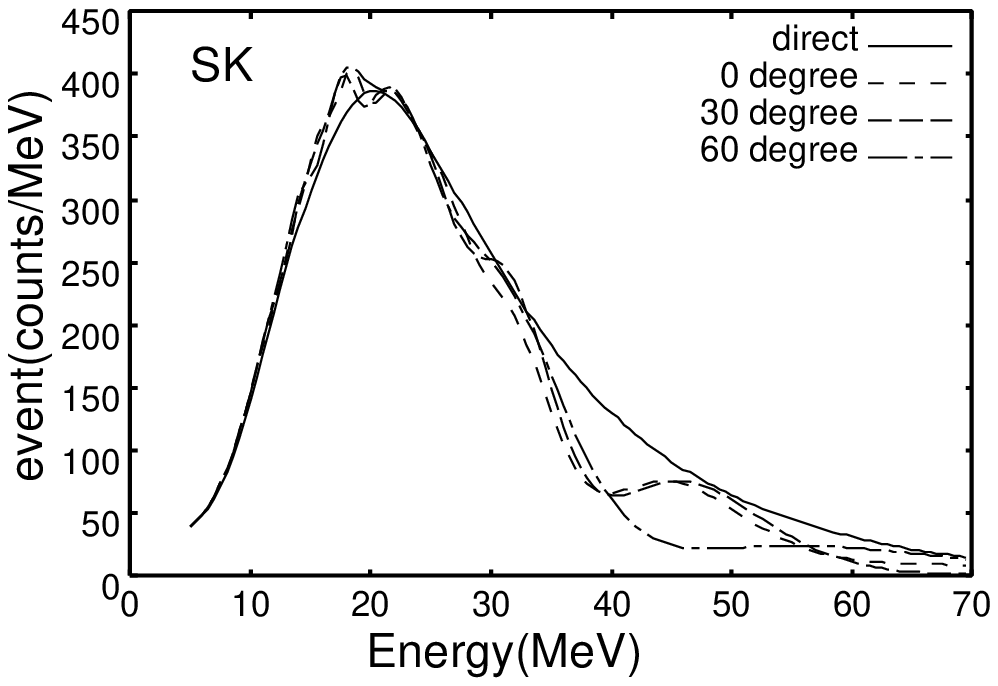}
\includegraphics[width=\hsize]{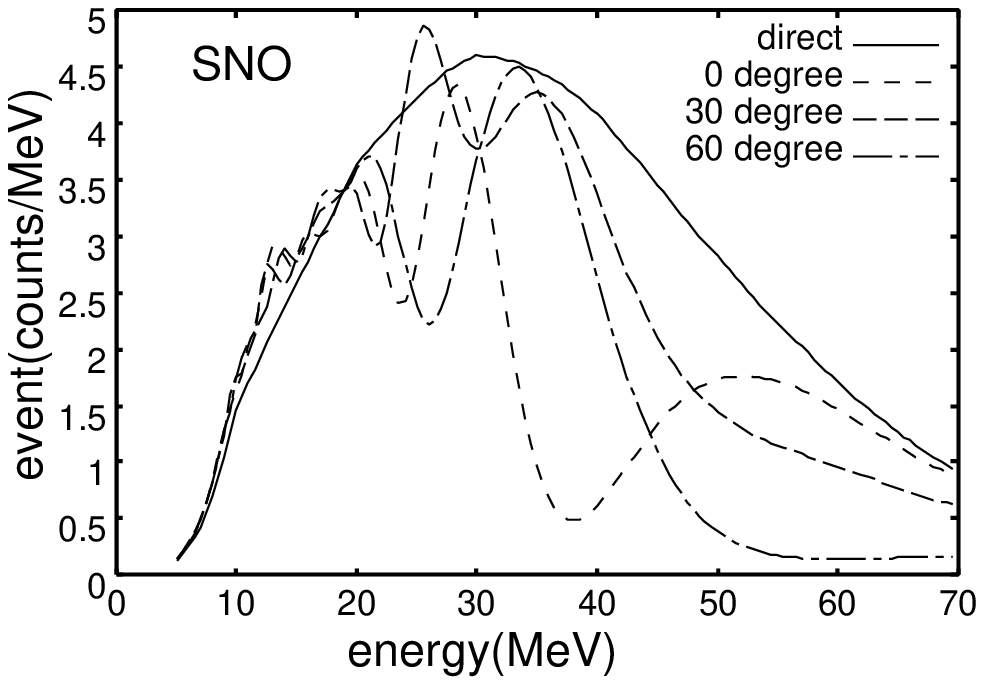}
\includegraphics[width=\hsize]{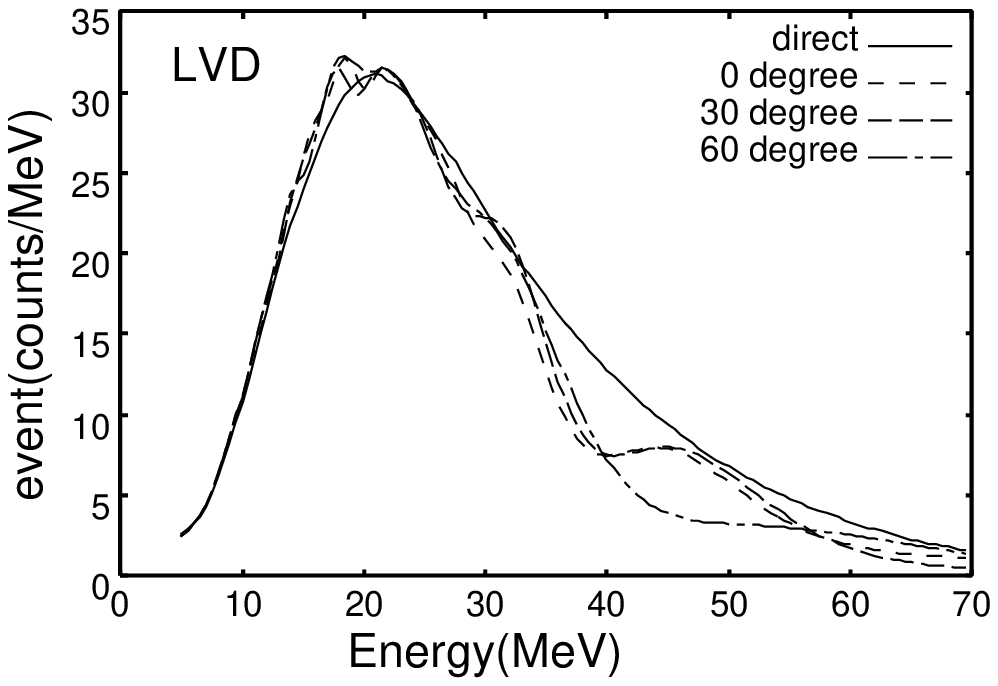}
\caption{Earth effect on SN spectra in different detectors as a
function of the nadir angle.  For SNO only CC events are taken into
account.  The oscillation parameters correspond to the solar LMA case
and a small $\theta_{13}$ as described in the text.  (Figures from
Ref.~\cite{Takahashi:2001dc} with permission.)}
\label{fig:4}
\end{figure}

Assuming that the SN~1987A neutrino anomalies are caused by
statistical flukes, we may gauge our expectations for a future SN by
theoretical predictions based on numerical simulations. In any case,
the signal of a future SN itself will determine if SN theory is
correct with regard to the neutrino fluxes and spectra. Taking the
numerical model of Fig.~\ref{fig:2} for the source, the
time-integrated spectra at Super-Kamiokande, SNO and LVD are shown in
Fig.~\ref{fig:4} as a function of the nadir angle which determines the
Earth-segment of the neutrino path. The oscillation parameters were
chosen for the LMA case with $\Delta m_{12}^2=2\times10^{-5}~{\rm
eV}^2$, $\Delta m_{13}^2=3.2\times10^{-3}~{\rm eV}^2$,
$\sin^2\theta_{12}=0.87$, and $\sin^2\theta_{23}=1.0$. The unknown
third mixing angle was chosen small as
$\sin^2\theta_{13}=1.0\times10^{-6}$. Figure~\ref{fig:4} illustrates
that rather dramatic modifications of the spectra can be expected for
certain cases.

It is difficult to anticipate everything about future data. If a
galactic SN is observed, what we can learn about neutrino oscillations
depends on the detectors operating at that time and their geographical
location. It will also depend on the true source properties regarding
flavor-dependent spectra and fluxes, and what is already known about
the neutrino oscillation parameters as input information at that time.
Many authors have studied these questions
\cite{Takahashi:2001dc,Chiu:1999fh,Dighe:1999bi,Dutta:2001nf,%
Fuller:1998kb,Lunardini:2001pb,Minakata:2000rx,Takahashi:2001ep,%
Takahashi:2000it,Barger:2001yx}.  It appears that one may well
distinguish between large and small values of the elusive
$\theta_{13}$ and to distinguish between normal or inverted mass
hierarchies, or even accurately pin down $\Delta m_{12}^2$.
Therefore, a SN neutrino observation would complement the upcoming
efforts of precision determination of neutrino oscillation parameters
in long-baseline
experiments~\cite{Barger:2000nf,Freund:2001ui,Cervera:2000kp}.

While neutrino oscillations are crucial for SN neutrino observations,
the smallness of the measured mass differences implies that
oscillations are {\it not\/} important in the SN core, and also not in
the SN atmosphere within the stalled shock wave because the
matter-induced weak potential dominates over the mass differences for
the flavor-dependent neutrino refractive index.  In these regions and
on the relevant time-scales the separate flavor lepton numbers are
effectively conserved, in spite of maximal neutrino
mixing~\cite{Wolfenstein:1979ni,Raffelt:bs,Hannestad:1999zy}.


\section{\uppercase{Sterile Neutrinos}}

The existence of sterile neutrino degrees of freedom is a logical
possibility that has received much attention. One possible role for
such particles is to constitute the cosmic dark matter.  Depending on
their masses and their mixings with active neutrinos, they can be hot,
warm, or cold dark matter~\cite{Abazajian:2001nj}.  These particles
would be emitted from SN cores so that the SN~1987A energy-loss
argument (Sec.~\ref{sec:LimitsonNewParticles}) provides some of the
most important constraints on this conjecture.

If sterile neutrinos have masses in the eV range and mix with active
flavors, they can modify the nucleosynthesis processes that take place
in the neutrino-driven wind of a SN core after the explosion. The
r-process nucleosynthesis of heavy elements requires a neutron-rich
environment. The $n/p$ ratio is governed by the $\beta$ processes
involving the $\nu_e$ and $\bar\nu_e$ flux.  In standard SN
calculations the required conditions for r-process nucleosynthesis are
not achieved. However, if $\nu_e\to\nu_{\rm sterile}$ oscillations are
efficient enough, the $\nu_e+n\to p+e^-$ reactions are quenched,
reducing the proton fraction, and thus allowing enough neutrons to
escape being trapped in $\alpha$
particles~\cite{Caldwell:1999zk,Patel:1999hm}.  Therefore, low-mass
sterile neutrinos can play a crucial role in this environment.


\section{\uppercase{Neutrino Mass Sensitivity}}

The ever accumulating evidence for neutrino oscillations and for
neutrino mass differences in the 50~meV range and below has reduced
the neutrino mass question to one unknown overall mass scale $m_\nu$
that could be much larger than the mass differences. Tritium end point
experiments reveal $m_\nu<2.8~{\rm eV}$
\cite{Weinheimer:tn,Lobashev:tp}, the future sensitivity at KATRIN may
reach the $0.3~{\rm eV}$ level~\cite{Osipowicz:2001sq}.  The observed
power spectrum of the galaxy distribution function and of the cosmic
microwave background radiation reveals similar constraints and future
sensitivities~\cite{Tegmark:2000qy}.

The SN~1987A signal duration gave a time-of-flight limit of
$m_\nu\alt20~{\rm eV}$~\cite{Loredo:mk}, a refined recent analysis
even claims $m_\nu<6~{\rm eV}$ at 95\%~CL~\cite{Loredo:2001rx}. The
neutrino signal of a galactic SN observed in Super-Kamiokande would be
sensitive to about 3~eV \cite{Totani:1998nf}.  If a black hole forms a
few seconds after the original collapse, the quick termination of the
neutrino burst imprints a structure on the neutrino light curve,
corresponding to an $m_\nu$ dispersion sensitivity of about
2~eV~\cite{Beacom:2000qy}.  A further improvement is possible if the
SN collapse emits a measurable gravitational wave signal that can
serve as a zero-point for the neutrino time-of-flight delay.
Independently of black-hole formation, a Super-Kamiokande mass
sensitivity of around 1~eV has been claimed~\cite{Arnaud:2001gt}.
Further improvements with a megatonne detector may be possible, but
have not been investigated in detail.

Therefore, while the SN time-of-flight method would provide new and
independent direct limits on the neutrino mass, this method does not
seem competitive with future tritium endpoint and cosmological
sensitivities. None of these methods seems able to reach the crucial
50~meV range characteristic of the neutrino mass differences.


\section{\uppercase{Limits on New Particles}}

\label{sec:LimitsonNewParticles}

The neutrino signal of SN~1987A has been used to derive numerous
limits on new particles or novel neutrino properties. One standard
argument holds that the signal duration of about 10~s precludes that
too much energy was carried away by axions, right-handed (sterile)
neutrinos or other exotic channels. This classic ``energy-loss
argument'' has been applied to constrain axion or Majoron
interactions, neutrino dipole moments, active-sterile mixings, or
right-handed currents~\cite{Raffelt:wa,Raffelt:1999tx}.  Most
recently, it has been used to constrain the compactification scale of
large extra dimensions by constraining the emission of Kaluza-Klein
gravitons~\cite{Cullen:1999hc,Hanhart:2001fx}.

The SN~1987A energy-loss argument is problematic because far-reaching
conclusions depend on a few late events in the Kamiokande~II and IMB
detectors. Evidently a high-statistics neutrino lightcurve from a
future galactic SN would place such limits on firm experimental
grounds.

Of course, not all SN particle-physics limits depend on the sparse
SN~1987A data. For SN graviton emission in theories with large extra
dimensions, the $\gamma$-rays from the subsequent decay of the
Kaluza-Klein gravitons can leave observable signatures in the cosmic
$\gamma$-ray background~\cite{Hannestad:2001jv} or from young SN
remnants and neutron stars~\cite{Hannestad:2001xi}.


\section{\uppercase{New Phases of Nuclear\\ Matter}}

Standard numerical SN simulations generally rely on a nuclear equation
of state and neutrino opacities which are based on the assumption that
nuclear matter at all relevant densities and temperatures is well
described in terms of nucleons. However, the QCD phase diagram in the
temperature-density plane may be far more complicated.  One
long-standing speculation holds that the true ground state of dense
matter consists of quarks rather than nucleons, leading to various
modifications of standard SN and neutron-star
physics~\cite{Madsen:1998uh,Weber:qn}. More recently, the existence of
an intruiging color-superconducting phase has been
debated~\cite{Alford:2001dt}.

When the equation of state and/or the neutrino opacities suddenly
change during the first few seconds after SN collapse due to a nuclear
phase transition, an observable signature in the neutrino light curve
could obtain. For example, instead of tapering off, the neutrino
luminosity could show a second burst~\cite{Carter:2000xf,Hong:2001gt}
or could suddenly terminate by a secondary collapse to a black
hole~\cite{Pons:2001ar}.  Evidently, a high-statistics neutrino light
curve from a galactic SN would shed new light on the existence or
non-existence of new phases of nuclear matter.


\section{\uppercase{Cosmic Relic Neutrinos\\ from past Supernovae}}

All supernovae which occurred since the birth of the universe
contribute to a cosmic background of neutrinos in the energy range up
to about 50~MeV. A simple estimate shows that the average neutrino
luminosity of a galaxy from stellar collapse is roughly comparable to
its optical photon luminosity. If the past SN rate is assumed to be
constant at the present-day levels of Table~\ref{tab:SNrates}, the SN
relic neutrinos amount to an approximate flux of $1~\rm
cm^{-2}~s^{-1}$. However, when galaxies first formed they must have
been much more active at star formation, leading to flux estimates of
\hbox{5--$50~\rm cm^{-2}~s^{-1}$}
\cite{Totani:1995dw,Hartmann:qe,Kaplinghat:1999xi} where the high
number is thought to be a plausible upper limit.  A positive detection
of this flux would provide a new window to the universe at redshifts
of a few.

Such flux levels are, in principle, detectable because they stick
above solar and atmospheric neutrinos for $20\alt E_\nu\alt 50~{\rm
MeV}$.  The limit from the Kamiokande~II detector is $226~\rm
cm^{-2}~s^{-1}$ (90\%~CL) for energies 19--35~MeV \cite{Zhang:tv}.  A
preliminary Super-Kamiokande limit is $39~\rm cm^{-2}~s^{-1}$
\cite{Totsuka2001}, assuming the energy spectrum
of~\cite{Kaplinghat:1999xi}, i.e.\ Super-Kamiokande has touched the
upper range of theoretical estimates.

A further improvement of the sensitivity requires a new detector
concept because Super-Kamiokande is limited by an irreducible
background of ``invisible muons,'' i.e.\ sub-Cherenkov muons produced
by atmospheric neutrinos. Their subsequent decays produce electrons or
positrons in the energy window of the SN relic neutrinos.


\section{\uppercase{Conclusions}}

Core-collapse supernovae are powerful neutrino sources. The
observation of the SN~1987A neutrino burst has provided a crude
confirmation of the idea that stellar collapse leads to a neutrino
burst which carries away the gravitational binding energy of the
collapsed object, but leaves many questions open.

The high-statistics neutrino observation of a galactic SN would allow
one to watch directly the stellar collapse, to confirm or refute the
delayed explosion mechanism, and to search for signatures of new
nuclear phases in the late-time behavior of the neutrino light
curve. The neutrino burst precedes the optical explosion by a few
hours, hence an early warning can be given to direct telescopes in the
SN direction in the sky.

Many of the classic SN~1987A particle-physics limits are problematic
because of the sparse data. A high-statistics observation would 
provide these important results with a firm experimental basis.

One would obtain new time-of-flight neutrino mass limits, but neutrino
masses in the sub-eV range will likely remain elusive.

The detailed characteristics of the neutrino signal can discriminate
between different neutrino mass and mixing scenarios. If the SN
neutrinos propagate through the Earth before reaching the detector,
spectacular regeneration effects can obtain in some cases. The
SN~1987A signal interpretation already requires including the Earth
effect if the ``solar-neutrino mixing angle'' is large.

In summary, the high-statistics neutrino observation of a future
galactic SN guarantees a rich astrophysical, particle-physics, and
nuclear-physics harvest. Of course, the SN rate is low, but still, the
neutrinos from about a thousand galactic SNe are on their
way. Hopefully one of these bursts will be intercepted at Earth by one
or more large neutrino observatories.


\section*{\uppercase{Acknowledgments}}

In Munich, this work was partly supported by the Deut\-sche
For\-schungs\-ge\-mein\-schaft under grant No.\ SFB 375 and the ESF
network Neutrino Astrophysics.



\begin{thebibliography}{99}

\bibitem{Brown}
G.~E.~Brown, H.~A.~Bethe and G.~Baym
``Supernova theory,''
Nucl. Phys. {\bf A375} (1982) 481.

\bibitem{Woosley:ta}
S.~E.~Woosley and T.~A.~Weaver,
``The physics of supernova explosions,''
Ann.\ Rev.\ Astron.\ Astrophys.\  {\bf 24} (1986) 205.

\bibitem{Petschek}
A.~G.~Petschek (ed.),
``Supernovae,''
Springer-Verlag, New York, 1990.

\bibitem{Burrows:mk}
A.~Burrows,
``Supernova explosions in the universe,''
Nature {\bf 403} (2000) 727.

\bibitem{Cappellaro}
E.~Cappellaro and M.~Turatto, 
``Supernova types and rates,'' 
astro-ph/0012455.

\bibitem{Riess:1998cb}
A.~G.~Riess {\it et al.}  [Supernova Search Team Collaboration],
``Observational evidence from supernovae for an accelerating 
universe and a cosmological constant,''
Astron.\ J.\  {\bf 116} (1998) 1009.

\bibitem{Perlmutter:1998np}
S.~Perlmutter {\it et al.}  
[Supernova Cosmology Project Collaboration],
``Measurements of Omega and Lambda from 42 high-redshift 
supernovae,''
Astrophys.\ J.\  {\bf 517} (1999) 565.

\bibitem{Cappellaro:1999qy}
E.~Cappellaro, R.~Evans and M.~Turatto,
``A new determination of supernova rates and a comparison with 
indicators for galactic star formation,''
Astron. Astrophys. {\bf 351} (1999) 459. 

\bibitem{Asiago}
The Asiago Supernova Catalogue,\\ 
http://merlino.pd.astro.it/$\sim$supern/

\bibitem{vandenBergh1991}
S.~van den Bergh and G.~A.~Tammann
``Galactic and extragalactic supernova rates''
Ann. Rev. Astron. Astrophys. 29 (1991) 363.

\bibitem{Tammann:ev}
G.~A.~Tammann, W.~L\"offler and A.~Schr\"oder,
``The galactic supernova rate,''
Astrophys.\ J.\ Supp.\  {\bf 92} (1994) 487.

\bibitem{Strom1994}
R.~G.~Strom,
``Guest stars, sample completeness and the local supernova rate''
Astron. Astrophys. 288 (1994) L1.

\bibitem{Alekseev:dy}
E.~N.~Alekseev {\it et al.},
``Upper bound on the collapse rate of massive stars in the Milky Way 
given by neutrino observations with the Baksan underground telescope,''
J.\ Exp.\ Theor.\ Phys.\  {\bf 77} (1993) 339
[Zh.\ Eksp.\ Teor.\ Fiz.\  {\bf 104} (1993) 2897].

\bibitem{Font}
J.~A.~Font, ``Numerical hydrodynamics in general relativity,''
Living Rev. Relativity {\bf 3} (2000) 2.
Online article: cited on 3 Dec 2001,
http://www.livingreviews.org/Articles /Volume3/2000-2font/ 

\bibitem{Rampp:2000ws}
M.~Rampp and H.~T.~Janka,
``Spherically symmetric simulation with Boltzmann neutrino transport 
of core collapse and post-bounce evolution of a 15 solar mass star,''
Astrophys.\ J.\  {\bf 539} (2000) L33.

\bibitem{Mezzacappa:2000jb}
A.~Mezzacappa, M.~Liebend\"orfer, O.~E.~Mes\-ser, W.~R.~Hix, 
F.~K.~Thielemann and S.~W. Bruenn,
 ``The simulation of a spherically symmetric supernova of a 13 solar 
mass star with Boltzmann neutrino transport, and its implications 
for the supernova mechanism,''
Phys.\ Rev.\ Lett.\  {\bf 86} (2001) 1935.

\bibitem{Burrows:tj}
A.~Burrows, T.~Young, P.~Pinto, R.~Eastman and T.~A.~Thompson,
``A new algorithm for supernova neutrino transport and some 
applications,''
Astrophys.\ J.\  {\bf 539} (2000) 865.

\bibitem{Bethe:1984ux} 
H.~A.~Bethe and J.~R.~Wilson, 
``Revival of a stalled supernova shock by neutrino heating,'' 
Astrophys.\ J.\ {\bf 295} (1985) 14.

\bibitem{Janka:2000bt}
H.~T.~Janka,
``Conditions for shock revival by neutrino heating in core-collapse  
supernovae,''
Astron. Astrophys. {\bf 368} (2001) 527.

\bibitem{Totani:1997vj}
T.~Totani, K.~Sato, H.~E.~Dalhed and J.~R.~Wilson,
``Future detection of supernova neutrino burst and 
explosion mechanism,''
Astrophys.\ J.\  {\bf 496} (1998) 216.

\bibitem{Wilson:rx}
J.~R.~Wilson and R.~W.~Mayle,
``Report on the progress of supernova research by the Livermore 
group,''
Phys.\ Rept.\  {\bf 227} (1993) 97.

\bibitem{Falk:kf}
S.~W.~Falk and D.~N.~Schramm,
``Limits from supernovae on neutrino radiative lifetimes,''
Phys.\ Lett.\ B {\bf 79} (1978) 511.

\bibitem{Schramm:1981mk}
D.~N.~Schramm and J.~R.~Wilson,
``Supernovae induced by axion like particles,''
Astrophys.\ J.\  {\bf 260} (1982) 868.

\bibitem{Berezhiani:1999qh}
Z.~Berezhiani and A.~Drago,
``Gamma ray bursts via emission of axion-like particles,''
Phys.\ Lett.\ B {\bf 473} (2000) 281.

\bibitem{Fuller}
G.~M.~Fuller, R.~Mayle, B.~S.~Meyer and J.~R.~Wilson, 
``Can a closure mass neutrino help solve the supernova shock 
reheating problem?''
Astrophys. J. {\bf 389} (1992) 517.

\bibitem{Lai}
D.~Lai, D.~F.~Chernoff and J.~M.~Cordes,
``Pulsar jets: Implications for neutron star kicks and initial 
spins,''
Astrophys. J. {\bf 549} (2001) 1111.

\bibitem{Raffelt:2001kv}
G.~G.~Raffelt,
``Mu- and tau-neutrino spectra formation in supernovae,''
Astrophys. J. 561 (2001) 890.

\bibitem{Janka:1995ir}
H.~T.~Janka, W.~Keil, G.~Raffelt and D.~Seckel,
``Nucleon spin fluctuations and the supernova emission of neutrinos 
and axions,''
Phys.\ Rev.\ Lett.\  {\bf 76} (1996) 2621.

\bibitem{Horowitz:2001xf}
C.~J.~Horowitz,
``Weak magnetism for antineutrinos in supernovae,''
arXiv:astro-ph/0109209.

\bibitem{Jegerlehner:1996kx}
B.~Jegerlehner, F.~Neubig and G.~Raffelt,
``Neutrino Oscillations and the Supernova 1987A Signal,''
Phys.\ Rev.\ D {\bf 54} (1996) 1194.

\bibitem{Hirata:ad}
K.~S.~Hirata {\it et al.},
``Observation in the Kamiokande-II detector of the neutrino burst 
from supernova SN 1987A,''
Phys.\ Rev.\ D {\bf 38} (1988) 448.

\bibitem{Bionta:1987qt}
R.~M.~Bionta {\it et al.},
``Observation of a neutrino burst in coincidence with supernova 
SN 1987A in the Large Magellanic Cloud,''
Phys.\ Rev.\ Lett.\  {\bf 58} (1987) 1494.

\bibitem{Alekseev:gp}
E.~N.~Alekseev, L.~N.~Alekseeva, I.~V.~Krivo\-sheina and 
V.~I.~Volchenko,
``Detection 
of the neutrino signal from SN 1987A in the 
LMC using the INR Baksan underground scintillation telescope,''
Phys.\ Lett.\ B {\bf 205} (1988) 209.

\bibitem{Virtue:2001mz}
C.~J.~Virtue  [SNO Collboration Collaboration],
``SNO and supernovae,''
Nucl.\ Phys.\ B Proc.\ Suppl.\  {\bf 100} (2001) 326.

\bibitem{Aglietta:2001jf}
M.~Aglietta {\it et al.},
``Effects of neutrino oscillations on the supernova signal in LVD,''
arXiv:astro-ph/0112312.

\bibitem{Piepke:tg}
A.~Piepke  [KamLAND Collaboration],
``Kamland: A reactor neutrino experiment testing the solar neutrino 
anomaly,''
Nucl. Phys. B (Proc.\ Suppl.) {\bf 91} (2001) 99.

\bibitem{Cadonati:2000kq}
L.~Cadonati, F.~P.~Calaprice and M.~C.~Chen,
``Supernova neutrino detection in Borexino,''
Astropart.\ Phys.\  {\bf 16} (2002) 361.

\bibitem{Ahrens:2001tz}
J.~Ahrens {\it et al.}  [AMANDA Collaboration],
``Search for supernova neutrino-bursts with the AMANDA detector,''
Astropart.\ Phys.\  {\bf 16} (2002) 345.

\bibitem{Beacom2001}
J.~Beacom,
``Supernova neutrino physics: Measuring the source properties,''
Talk presented at Physics Potential of Supernova II Neutrino
Detection,
15--16 February 2001, Marina del Rey, California,
http://www.physics.ucla.edu/SNII/abstracts /23.pdf

\bibitem{Vogel:2001yi}
P.~Vogel,
``Future galactic supernova neutrino signal: What can we learn?,''
arXiv:nucl-th/0111016.

\bibitem{Smith:td}
P.~F.~Smith,
``Omnis: An improved low cost detector to measure mass and mixing 
of mu tau neutrinos from a galactic supernova,''
Astropart.\ Phys.\  {\bf 8} (1997) 27.

\bibitem{Boyd2001}
R.~N.~Boyd and A.~S.~J.~Murphy [OMNIS collaboration],
``The observatory for multiflavor neutrinos from supernovae,''
Nucl. Phys. A {\bf 688} (2001) 17.

\bibitem{Shiozawa2001} 
M.~Shiozawa, 
``Hyper-Kamiokande project,''
Talk at the International Workshop
on a Next Generation Long-Baseline Neutrino Oscillation Experiment,
30--31 May 2001, Epochal Tsukuba, Japan, 
http://neutrino.kek.jp/ jhfnu/workshop2/ohp/shiozawa.pdf.

\bibitem{Vagins2001}
M.~Vagins,
``Supernova neutrino requirements of UNO,''
Talk at the Conference on Underground Science
4--7 October 2001, Lead, South Dakota,
http://mocha.phys.washington.edu/$\sim$int\_talk 
/NUSL/2001/People/Vagins\_M

\bibitem{Frejus2002}
Workshop on ``Large Detectors for Proton Decay, Supernovae and 
Atmospheric Neutrinos and Low Energy Neutrinos from High
Intensity Beams,'' 16--18 January 2002, CERN, Geneva,
http://muonstoragerings.cern.ch/
NuWorkshop02/welcome.html

\bibitem{SNEWS}
SNEWS: SuperNova Early Warning System
http://hep.bu.edu/~snnet/

\bibitem{Beacom:1998fj}
J.~F.~Beacom and P.~Vogel,
``Can a supernova be located by its neutrinos?,''
Phys.\ Rev.\ D {\bf 60} (1999) 033007.

\bibitem{Ando:2001zi}
S.~Ando and K.~Sato,
``Determining the supernova direction by its neutrinos,''
arXiv: hep-ph/0110187.

\bibitem{Smirnov:ku}
A.~Y.~Smirnov, D.~N.~Spergel and J.~N.~Bahcall,
``Is Large Lepton Mixing Excluded?,''
Phys.\ Rev.\ D {\bf 49} (1994) 1389.

\bibitem{Lunardini:2000sw}
C.~Lunardini and A.~Y.~Smirnov,
``Neutrinos from SN1987A, Earth matter effects and the LMA solution
of the solar neutrino problem,''
Phys.\ Rev.\ D {\bf 63} (2001) 073009.

\bibitem{Kachelriess:2001sg}
M.~Kachelriess, A.~Strumia, R.~Tomas and J.~W.~Valle,
``SN1987A and the status of oscillation solutions to the solar 
neutrino problem,''
arXiv:hep-ph/0108100.

\bibitem{Takahashi:2001dc}
K.~Takahashi and K.~Sato,
``Earth effects on supernova neutrinos and their implications for
neutrino parameters,''
arXiv:hep-ph/0110105.

\bibitem{Chiu:1999fh}
S.~H.~Chiu and T.~K.~Kuo,
``Effects of neutrino temperatures and mass hierarchies on the 
detection  of supernova neutrinos,''
Phys.\ Rev.\ D {\bf 61} (2000) 073015.

\bibitem{Dighe:1999bi}
A.~S.~Dighe and A.~Y.~Smirnov,
``Identifying the neutrino mass spectrum from the neutrino burst 
from a supernova,''
Phys.\ Rev.\ D {\bf 62} (2000) 033007.

\bibitem{Dutta:2001nf}
G.~Dutta, D.~Indumathi, M.~V.~Murthy and G.~Rajasekaran,
``Neutrinos from stellar collapse: Comparison of signatures in water
and heavy water detectors,''
Phys.\ Rev.\ D {\bf 64} (2001) 073011.

\bibitem{Fuller:1998kb}
G.~M.~Fuller, W.~C.~Haxton and G.~C. McLaughlin,
``Prospects for detecting supernova neutrino flavor oscillations,''
Phys.\ Rev.\ D {\bf 59} (1999) 085005.

\bibitem{Lunardini:2001pb}
C.~Lunardini and A.~Y.~Smirnov,
``Supernova neutrinos: Earth matter effects and neutrino mass 
spectrum,''
Nucl.\ Phys.\ B {\bf 616} (2001) 307.

\bibitem{Minakata:2000rx}
H.~Minakata and H.~Nunokawa,
``Inverted hierarchy of neutrino masses disfavored by 
supernova 1987A,''
Phys.\ Lett.\ B {\bf 504} (2001) 301.

\bibitem{Takahashi:2001ep}
K.~Takahashi, M.~Watanabe, K.~Sato and T.~Totani,
``Effects of neutrino oscillation on the supernova neutrino 
spectrum,''
Phys.\ Rev.\ D {\bf 64} (2001) 093004.

\bibitem{Takahashi:2000it}
K.~Takahashi, M.~Watanabe and K.~Sato,
``The earth effects on the supernova neutrino spectra,''
Phys.\ Lett.\ B {\bf 510} (2001) 189.

\bibitem{Barger:2001yx}
V.~Barger, D.~Marfatia and B.~P.~Wood,
``Inverting a supernova: Neutrino mixing, temperatures and binding
energy,''
arXiv:hep-ph/0112125.

\bibitem{Barger:2000nf}
V.~D.~Barger, S.~Geer, R.~Raja and K.~Whisnant,
``Exploring neutrino oscillations with superbeams,''
Phys.\ Rev.\ D {\bf 63} (2001) 113011.

\bibitem{Freund:2001ui}
M.~Freund, P.~Huber and M.~Lindner,
``Systematic exploration of the neutrino factory parameter space
including errors and correlations,''
Nucl.\ Phys.\ B {\bf 615} (2001) 331.

\bibitem{Cervera:2000kp}
A.~Cervera, A.~Donini, M.~B.~Gavela, J.~J.~Gomez Cadenas, 
P.~Hernandez, O.~Mena and S.~Rigolin,
``Golden measurements at a neutrino factory,''
Nucl.\ Phys.\ B {\bf 579} (2000) 17
[Erratum-ibid.\ B {\bf 593} (2000) 731].

\bibitem{Wolfenstein:1979ni}
L.~Wolfenstein,
``Neutrino sscillations and stellar collapse,''
Phys.\ Rev.\ D {\bf 20} (1979) 2634.

\bibitem{Raffelt:bs}
G.~Raffelt and G.~Sigl,
``Neutrino flavor conversion in a supernova core,''
Astropart.\ Phys.\  {\bf 1} (1993) 165.

\bibitem{Hannestad:1999zy}
S.~Hannestad, H.~T.~Janka, G.~G.~Raffelt and G.~Sigl,
``Electron-, mu-, and tau-number conservation in a supernova core,''
Phys.\ Rev.\ D {\bf 62} (2000) 093021.

\bibitem{Abazajian:2001nj}
K.~Abazajian, G.~M.~Fuller and M.~Patel,
``Sterile neutrino hot, warm, and cold dark matter,''
Phys.\ Rev.\ D {\bf 64} (2001) 023501.

\bibitem{Caldwell:1999zk}
D.~O.~Caldwell, G.~M.~Fuller and Y.~Z.~Qian,
``Sterile neutrinos and supernova nucleosynthesis,''
Phys.\ Rev.\ D {\bf 61} (2000) 123005.

\bibitem{Patel:1999hm}
M.~Patel and G.~M.~Fuller,
``What are sterile neutrinos good for?,''
arXiv:hep-ph/0003034.


\bibitem{Weinheimer:tn}
C.~Weinheimer {\it et al.},
``High precision measurement of the tritium beta spectrum near
its  endpoint and upper limit on the neutrino mass,''
Phys.\ Lett.\ B {\bf 460} (1999) 219.

\bibitem{Lobashev:tp}
V.~M.~Lobashev {\it et al.},
``Direct search for mass of neutrino and anomaly in the tritium
beta-spectrum,''
Phys.\ Lett.\ B {\bf 460} (1999) 227.

\bibitem{Osipowicz:2001sq}
A.~Osipowicz {\it et al.}  [KATRIN Collaboration],
``KATRIN: A next generation tritium beta decay experiment with sub-eV
sensitivity for the electron neutrino mass,''
arXiv:hep-ex/0109033.

\bibitem{Tegmark:2000qy}
M.~Tegmark, M.~Zaldarriaga and A.~J.~Hamilton,
``Towards a refined cosmic concordance model: Joint 11-parameter 
constraints from CMB and large-scale structure,''
Phys.\ Rev.\ D {\bf 63} (2001) 043007.

\bibitem{Loredo:mk}
T.~J.~Loredo and D.~Q.~Lamb,
``Neutrinos from SN 1987A: Implications for cooling of the nascent
neutron star and the mass of the electron anti-neutrino,''
in Proc.~Fourteenth Texas Symposium on Relativistic Astrophysics,
Ann. N.Y. Acad. Sci. {\bf 571} (1989) 601.

\bibitem{Loredo:2001rx}
T.~J.~Loredo and D.~Q.~Lamb,
``Bayesian analysis of neutrinos observed from supernova SN 1987A,''
arXiv:astro-ph/0107260.

\bibitem{Totani:1998nf}
T.~Totani,
``Electron neutrino mass measurement by supernova neutrino bursts 
and implications on hot dark matter,''
Phys.\ Rev.\ Lett.\  {\bf 80} (1998) 2039.

\bibitem{Beacom:2000qy}
J.~F.~Beacom, R.~N.~Boyd and A.~Mezzacappa,
``Black hole formation in core collapse supernovae and time-of-flight
measurements of the neutrino masses,''
Phys.\ Rev.\ D {\bf 63} (2001) 073011.

\bibitem{Arnaud:2001gt}
N.~Arnaud, M.~Barsuglia, M.~A.~Bizouard, F.~Cavalier, M.~Davier, 
P.~Hello and T.~Pradier,
``Gravity wave and neutrino bursts from stellar collapse: 
A sensitive  test of neutrino masses,''
arXiv:hep-ph/0109027.

\bibitem{Raffelt:wa}
G.~G.~Raffelt,
``Stars as laboratories for fundamental physics,''
{\it  Chicago, USA: University Press (1996)}.

\bibitem{Raffelt:1999tx}
G.~G.~Raffelt,
``Particle physics from stars,''
Ann.\ Rev.\ Nucl.\ Part.\ Sci.\  {\bf 49} (1999) 163.

\bibitem{Cullen:1999hc}
S.~Cullen and M.~Perelstein,
``SN1987A constraints on large compact dimensions,''
Phys.\ Rev.\ Lett.\  {\bf 83} (1999) 268.

\bibitem{Hanhart:2001fx}
C.~Hanhart, J.~A.~Pons, D.~R.~Phillips and S.~Reddy,
``The likelihood of GODs' existence: Improving the SN 1987a constraint 
on the size of large compact dimensions,''
Phys.\ Lett.\ B {\bf 509} (2001) 1.

\bibitem{Hannestad:2001jv}
S.~Hannestad and G.~Raffelt,
``New supernova limit on large extra dimensions,''
Phys.\ Rev.\ Lett.\  {\bf 87} (2001) 051301.

\bibitem{Hannestad:2001xi}
S.~Hannestad and G.~G.~Raffelt,
``Stringent neutron-star limits on large extra dimensions,''
arXiv:hep-ph/0110067.

\bibitem{Madsen:1998uh}
J.~Madsen,
``Physics and astrophysics of strange quark matter,''
arXiv:astro-ph/9809032.

\bibitem{Weber:qn}
F.~Weber,
``Pulsars as astrophysical laboratories for nuclear and particle
physics,''
{\it  Bristol, UK: IOP (1999)}.

\bibitem{Alford:2001dt}
M.~G.~Alford,
``Color superconducting quark matter,''
Annu. Rev. Nucl. Part. Sci. 51 (2001) 131.

\bibitem{Carter:2000xf}
G.~W.~Carter and S.~Reddy,
``Neutrino propagation in color superconducting quark matter,''
Phys.\ Rev.\ D {\bf 62} (2000) 103002.

\newpage

\bibitem{Hong:2001gt}
D.~K.~Hong, S.~D.~Hsu and F.~Sannino,
``Supernovae, hypernovae and color superconductivity,''
Phys.\ Lett.\ B {\bf 516} (2001) 362.

\bibitem{Pons:2001ar}
J.~A.~Pons, A.~W.~Steiner, M.~Prakash and J.~M.~Lattimer,
``Evolution of proto-neutron stars with quarks,''
Phys.\ Rev.\ Lett.\  {\bf 86} (2001) 5223.

\bibitem{Totani:1995dw}
T.~Totani, K.~Sato and Y.~Yoshii,
``Spectrum of the supernova relic neutrino background and evolution 
of galaxies,''
Astrophys.\ J.\  {\bf 460} (1996) 303.

\bibitem{Hartmann:qe}
D.~H.~Hartmann and S.~E.~Woosley,
``The cosmic supernova neutrino background,''
Astropart.\ Phys.\  {\bf 7} (1997) 137.

\bibitem{Kaplinghat:1999xi}
M.~Kaplinghat, G.~Steigman and T.~P.~Walker,
``The supernova relic neutrino background,''
Phys.\ Rev.\ D {\bf 62} (2000) 043001.

\bibitem{Zhang:tv}
W.~Zhang {\it et al.},
``Experimental limit on the flux of relic anti-neutrinos 
from past supernovae,''
Phys.\ Rev.\ Lett.\  {\bf 61} (1988) 385.

\bibitem{Totsuka2001}
Y.~Totsuka, private communication (2001).

\end{thebibliography}
\end{document}